\documentclass[prb,twocolumn,epsfig,epsf]{revtex4}
\usepackage{amsmath,amsthm,amsfonts,amscd,mathrsfs,pifont,bm}
\usepackage{eucal,graphicx,color}
\usepackage{esint}

\makeatletter
\@for\next:={int,iint,iiint,iiiint,dotsint,oint,oiint,sqint,sqiint,
  ointctrclockwise,ointclockwise,varointclockwise,varointctrclockwise,
  fint,varoiint,landupint,landdownint}\do{%
    \expandafter\edef\csname\next\endcsname{%
      \noexpand\DOTSI
      \expandafter\noexpand\csname\next op\endcsname
      \noexpand\ilimits@
    }%
  }
\makeatother

\newcommand{\beq}{\begin{equation}}
\newcommand{\eeq}{\end{equation}}

\renewcommand{\Re}{\mathrm{Re}}
\renewcommand{\Im}{\mathrm{Im}}

\newcommand{\CC}{{\mathbb C}}
\newcommand{\RR}{{\mathbb R}}
\newcommand{\ZZ}{{\mathbb Z}}

\newcommand{\M}[1]{\pmb{#1}}

\begin{document}

\title  {Which Metrics Are Consistent with a Given Pseudo-Hermitian Matrix?}

\author {Joshua Feinberg\footnote{ https://orcid.org/0000-0002-2869-0010}}

\affiliation
       {Department of Physics\\ and\\
Haifa Research Center for Theoretical Physics and Astrophysics University of Haifa, Haifa 31905, Israel}

\author {Miloslav Znojil }

\affiliation
       {The Czech Academy of Sciences, Nuclear Physics Institute, Hlavn\'{\i} 130,
250 68 \v{R}e\v{z}, Czech Republic\\ and\\
Institute of System Science, Durban University of Technology, P.O.Box 1334, Durban, 4000, South Africa}

\date   {September 22, 2021}

\begin  {abstract}
Given a diagonalizable $N\times N$ matrix $\M{H}$, whose non-degenerate spectrum consists of $p$ pairs of complex conjugate eigenvalues and additional $N-2p$ real eigenvalues, we determine all metrics $\M{M}$, of all possible signatures, with respect to which $\M{H}$ is pseudo-hermitian. In particular, we show that any compatible $\M{M}$ must have $p$ pairs of opposite eigenvalues in its spectrum so that $p$ is the minimal number of both positive and negative eigenvalues of $\M{M}$. We provide explicit parametrization of the space of all admissible metrics and show that it is topologically a $p$-dimensional torus tensored with an appropriate power of the group $\ZZ_2$.  
\end{abstract}

\maketitle
\newpage
\section{Introduction} 
There is intensive ongoing interest in PT-symmetric quantum mechanics (PTQM) \cite{BB} and its broader applications (see \cite{BGSZ, Christod, CMB} for recent reviews). The Hilbert space of a PT-symmetric quantum mechanical model is endowed with a non-trivial metric operator, with respect to which the hamiltonian is hermitian. Hamiltonians of PTQM models with proper metrics are sometimes referred to as {\it quasi-hermitian} \cite{Dieudonne, stellenbosch}. They are diagonalizable, and their spectrum is real, because they are similar to conventionally hermitian hamiltonians. 

In contrast, hamiltonians of PTQM models with broken $PT$-symmetry are associated with {\it indefinite} metrics, and  are referred to as {\it pseudo-hermitian} \cite{Froissart, Mostafazadeh, Mostafazadeh1, Mostafazadeh2}. They act on state vectors in Krein (or Pontryagin) space \cite{AK, Krein}. Their eigenvalues are either real or come in  complex-conjugate pairs \cite{BB, CMB, Zn, Zn1, Levai}. 

For a mathematically precise summary of the nomenclature of quasi-hermiticity and pseudo-hermiticity see  \cite{Fring-Assis}. 

Quasi-hermitian linear operators in finite dimensional space are represented by quasi-hermitian (QH) matrices. Thus, we say that the $N\times N$ matrix $\M{H}$ (the ``hamiltonian") is quasi-hermitian with respect to the $N\times N$ hermitian proper (non-negative) metric $\M{M}$, if $\M{H}$ and its adjoint fulfill the {\em intertwining relation}
\begin{equation}\label{intertwining}
\M{H}^\dagger \M{M} = \M{M}\M{H} \,.
\end{equation} 
(Note that both sides of this equation are hermitian.) The relation \eqref{intertwining} simply means that $\M{H}$ is hermitian in a vector space endowed with a non-trivial metric $\M{M}$, namely, 
\begin{equation}\label{metric}
\langle {\bf A}_1 | \M{H}{\bf A}_2\rangle_M  =  \langle \M{H}{\bf A}_1 | {\bf A}_2\rangle_M\,, 
\end{equation}
with inner product $\langle {\bf A}_1 | {\bf A}_2\rangle_M = \langle {\bf A}_1 | \M{M}{\bf A}_2\rangle$ (and where, of course, $\langle {\bf A}_1 | {\bf A}_2\rangle$ is the standard inner product, corresponding to $\M{M}=\M{1}$, with respect to which the adjoint $\M{H}^\dagger$ in (\ref{intertwining}) is defined).

{\it Strictly}-quasi-hermitian (SQH) matrices were introduced in \cite{FRDecember, MK, FR-review}. SQH matrices are hermitian with respect to positive definite (and therefore invertible) metrics $\M{M}$. This implies that $\M{H}$ is similar to a hermitian matrix, with the similarity matrix being essentially the (invertible) square root of $\M{M}$. Consequently, SQH are diagonalizable and have purely real eigenvalue spectrum.  In contrast, merely quasi-hermitian matrices are associated with non-negative possibly non-invertible metrics.

SQH random matrices arise, for example, in studying the spectrum of stable small oscillations of large mechanical systems with large connectivity \cite{FRDecember, MK}. The low-frequency part of the spectrum of such systems was found in \cite{FRDecember, MK} to be universal, with the phonon density of states tending to a non-zero constant at zero frequency. In other words, phonons in such systems have universal spectral dimension $d_s = 1$.

An interesting SQH random matrix model was introduced in \cite{JK}. These authors fixed a metric $\M{M}$, and took the hamiltonian $\M{H}$ as random, with the aim of studying numerically the dependence of the average density of eigenvalues and level spacing statistics on the metric. Yet another interesting example of an SQH random matrix model, akin to the Dicke model of superradiance, was provided by \cite{DGK}, in which a numerical study of the level spacing distribution was carried out. 

Upon truncation to finite vector spaces, pseudo-hermitian operators turn into pseudo-hermitian (PH) matrices. Such matrices are still defined by the intertwining relation \eqref{intertwining}, but now the metric $\M{M}$ has indefinite signature. A PH matrix, in contrast fo an SQH matrix, need not\footnote{A PH matrix $\M{H}$ may still have purely real eigenvalues. For example, $\M{H}=1\!\!1$ is PH with respect to a metric $\M{M}$ of any signature. A real diagonal $\M{H}$ and a diagonal $\M{M}$ of any signature is yet another example.} be similar to a hermitian matrix. In this case, since according to \eqref{intertwining}  $\M{H}$ is similar to its hermitian adjoint, it follows that the characteristic polynomial of $\M{H}$ has real coefficients: $\left(\det (z-\M{H})\right)^* = \det (z^*-\M{H}).$ Thus, as was asserted above concerning PH operators, the eigenvalues of $\M{H}$ are either real, or come in complex-conjugate pairs. 

Recently, a family of pseudo-hermitian random matrix models was introduced and reviewed in \cite{FR-review} and elaborated on in detail in \cite{FR-B}. The spectrum of these matrices, in the large-$N$ limit, typically occupies two two-dimensional islands in the complex plane, symmetric with respect to the real axis, together with a macroscopic fraction of real eigenvalues, and exhibits interesting structural ``phase transitions" as some parameters are changed. Finally, see \cite{KA} for a discussion of real asymmetric PH random matrices.

A generic $N\times N$ PH matrix $\M{H}$ is diagonalizable
\begin{equation}\label{generic}
\M{H} = \M{S}^{-1}\M{\Lambda}\M{S},\quad \M{S}\in GL(N,\CC)
\end{equation}
with non-degenerate spectrum 
\begin{equation}\label{spectrum}
\M{\Lambda} = {\rm diag}(\lambda_1,\ldots ,\lambda_r, z_1, z_1^*, \ldots, z_p, z_p^*),\,\, \lambda_i\in\RR, z_j\in\CC
\end{equation}
consisting of distinct $r$ real eigenvalues and $p$ pairs of complex conjugate eigenvalues, such that 
\begin{equation}\label{counting}
r+2p = N.
\end{equation} 
We can always choose $\M{S}$ such that the eigenvalues are ordered according to \eqref{spectrum}. Note further that $\M{S}$ is defined up to a multiplication on the left by an invertible  diagonal complex matrix 
\begin{equation}\label{D}
\M{D} = {\rm diag}(d_1,\ldots d_N),\quad \det \M{D}\neq 0,
\end{equation}
that is, $\M{S}$ and $\M{D} \M{S}$ lead to the same $\M{H}$, so in effect $\M{S}\in GL(N,\CC)/{\mathcal D}$, where ${\mathcal D}\subset GL(N,\CC)$ is the subgroup of invertible diagonal matrices. 
 
The $N$ eigenvectors of $\M{H}$ on the right and on the left are therefore, respectively,  
\begin{eqnarray}\label{eigenvectors}
\M{R}_\mu &=& \M{S}^{-1}\hat{\M{e}}_\mu\nonumber\\
\M{L}_\mu &=& \M{S}^\dagger\,\hat{\M{e}}_\mu,\quad  \mu:=1,\ldots, N
\end{eqnarray}
where $\hat{\M{e}}_\mu$ is the $\mu$-th Cartesian unit vector, and they comprise a complete biorthogonal basis.  Thus, changing $\M{S}\rightarrow\M{D}\M{S}$ has the effect of rescaling 
$\M{R}_\mu\rightarrow d_\mu^{-1}\M{R}_\mu$ and $\M{L}_\mu\rightarrow d_\mu^*\M{L}_\mu$, which does not affect completeness nor biorthogonality of the basis vectors. 

In this paper we consider \eqref{intertwining} as a linear equation for determining the metric $\M{M}$, given a generic diagonalizable PH matrix $\M{H}$ with non-degenerate spectrum \eqref{spectrum}. That is, given \eqref{generic}, our aim is to find all hermitian matrices $\M{M}$, of all possible signatures, which fulfill the intertwining relation \eqref{intertwining}.

The rest of this paper is organized as follows: In Section \ref{sec:solution}  we derive the general solution of \eqref{intertwining} for $\M{M}$, which depends on $r$ real parameters and $p$ complex ones. In Section \ref{sec:canonical} we disentangle the redundant degrees of freedom in that solution from the significant ones, and bring the resulting metric to its canonical form, which depends only on its inertia and an additional set of $p$ phases. We also specify the topology of the resulting space of solutions of \eqref{intertwining}. We conclude with a brief discussion of our results in Section \ref{Discussion}. \\

\section{Solving for $\M{M}$}\label{sec:solution}
We shall concentrate on finding invertible solutions $\M{M}$ to \eqref{intertwining}. The problem at hand clearly simplifies upon transforming to the diagonalizing basis \eqref{eigenvectors} of $\M{H}$. To this end, we parametrize the unknown metric $\M{M}$ by another hermitian matrix $\M{m}$ as
\begin{equation}\label{Mm}
\M{M} = \M{S}^\dagger\M{m}\M{S}.
\end{equation}
Thus, $\M{m}$ and $\M{M}$ are congruent, and by Sylvester's Law of Inertia\cite{Horn}, these two matrices have the same numbers of positive and negative (and by assumption, no zero) eigenvalues.  
By substituting \eqref{Mm} in \eqref{intertwining} we obtain 
\begin{equation}\label{intertwining1}
\M{\Lambda}^\dagger\M{m} = \M{m}\M{\Lambda},
\end{equation}
or equivalently, by taking the $\mu\nu$ matrix element of \eqref{intertwining1},
\begin{equation}\label{elements}
(\Lambda_\mu^*-\Lambda_\nu) m_{\mu\nu} = 0, \quad \mu,\nu :=1,\ldots,N.
\end{equation}
(Cf. Eq.~(8) in \cite{Mostafazadeh}.) According to \eqref{spectrum}, the set of $N$ indices is split naturally into a subset of indices $R=\{1,2,\ldots,r\}$ corresponding to the real eigenvalues, and a subset of indices $C=\{r+1, r+2,\ldots,N\}$ corresponding to the complex eigenvalues. This induces a natural splitting of $\M{m}$ into four blocks as 
\begin{equation}\label{blocks}
\M{m} = \left(\begin{array}{cc}\M{m}_{RR} & \M{m}_{RC}\\ \M{m}_{RC}^\dagger & \M{m}_{CC}\end{array}\right)
\end{equation}
in self-evident notation, with hermitian diagonal blocks. 

Upon substituting in \eqref{elements} two indices $\mu,\nu\in R$ we obtain $(\Lambda_\mu-\Lambda_\nu)m_{RR\mu \nu}=0$. Based on non-degeneracy of the spectrum \eqref{spectrum} we conclude that $m_{RR\mu\nu} = 0$, unless $\mu = \nu$. Therefore 
\begin{equation}\label{mRR}
\M{m}_{RR} = {\rm diag}(\mu_1,\ldots, \mu_r),\quad \mu_i\in\RR
\end{equation} 
must be diagonal, with real eigenvalues $\mu_1,\ldots,\mu_r$.  Upon substituting a pair of indices belonging to two different subsets we immediately conclude that 
\begin{equation}\label{mRC}
\M{m}_{RC} = 0.
\end{equation} 
Finally, upon substituting two indices $\mu,\nu\in C$ it is straightforward to see from \eqref{elements}, based on the non-degeneracy of the spectrum, that $m_{CC\mu\nu}$=0, unless $\mu$ and $\nu$ correspond to a pair of complex conjugate eigenvalues, that is, one of the indices, say $\mu = r + 2s +1$ and $\nu = r + 2s +2$, with $0\leq s\leq p-1$ (or the other way around). In this case, the $2\times 2$ block of $\M{m}_{CC}$ associated with these two indices must be of the form\footnote{As a simple example, with the indefinite metric $\M{m}_{CC}$ taken to be the parity operator, see Eq. (13) of \cite{Zn}.} 
\begin{equation}\label{mCC}
m_{CC\mu\nu} =  \left(\begin{array}{cc} 0 & \tau_s^* \\ \tau_s & 0\end{array}\right) = \Re\tau_s \sigma_x + \Im\tau_s \sigma_y, \quad \tau_s\in\CC,
\end{equation}
where $\sigma_x, \sigma_y$ are Pauli matrices. For notational convenience, let us introduce the planar vector $\M{\tau}_s  = \hat{\M{x}}\Re\tau_s + \hat{\M{y}}\Im\tau_s$ and write the matrix in \eqref{mCC} as $\M{\tau}_s\!\cdot\!\M{\sigma}$, where $\M{\sigma} = (\sigma_x,\sigma_y,\sigma_z)$. Let us now gather the results \eqref{mRR}-\eqref{mCC} and express our solution of \eqref{intertwining} as
\begin{equation}\label{m-solution}
\M{m} = {\rm block~diag}(\mu_1,\mu_2,\ldots,\mu_r,\M{\tau}_1\!\cdot\!\M{\sigma},\M{\tau}_2\!\cdot\!\M{\sigma},\ldots,\M{\tau}_p\!\cdot\!\M{\sigma}).
\end{equation}
Thus, our solution for $\M{m}$ is in fact a family of metrics, parametrized by the $r$ real parameters $\mu_i$ and the $p$ complex parameters $\tau_s$.

\section{Diagonalization of $\M{m}$ and Canonical Form of the Metric}\label{sec:canonical}
The $2\times 2$ block   $\M{\tau}_s\!\cdot\!\M{\sigma}$ in $\M{m}$ has eigenvalues $\pm|\tau_s|$. It can be diagonalized by first rotating it in the $xy$ plane onto the $x$-axis, followed by a rotation around the $y$-axis onto the $z$-direction:
\begin{eqnarray}\label{rotation}
&&U_s e^{i{n_s\pi\over 2}\sigma_z}\M{\tau}_s\!\cdot\!\M{\sigma} e^{-i{n_s\pi\over 2}\sigma_z} U_s^\dagger =(-1)^{n_s}|\tau_s| \sigma_z\nonumber\\
&&U_s = e^{i{\pi\over 4}\sigma_y}  e^{i{\arg\tau_s \over 2}\sigma_z}
\end{eqnarray}
with $n_s\in \{0,1\}$ chosen at will for each such block. Let us now pack the $p$ unitary blocks $U_s e^{i{n_s\pi\over 2}\sigma_z}$ into the unitary matrix 
\begin{equation}\label{unitary'}
{\mathcal U'} = {\rm block~diag}(\underbrace{1,\ldots,1}_{\text{r}},U_1 e^{i{n_1\pi\over 2}\sigma_z},\ldots,U_p e^{i{n_p\pi\over 2}\sigma_z}),
\end{equation}
which fully diagonalizes $\M{m}$ in \eqref{m-solution}:
\begin{eqnarray}\label{m-diagonal}
&&\hspace{-3mm}\M{m}_d = {\mathcal U'}\M{m} {\mathcal U'}^\dagger = \nonumber\\ 
&&\hspace{-3mm}{\rm block~diag}\left(\mu_1,\mu_2,\ldots,\mu_r,(-1)^{n_1}|\tau_1|\sigma_z,\ldots,(-1)^{n_p}|\tau_p|\sigma_z\right).\nonumber\\
\end{eqnarray}
It would be constructive to display the inertia of $\M{m}$ explicitly by rewriting the RHS of \eqref{m-diagonal} as 
\begin{eqnarray}\label{m-inertia}
\M{m}_d &=& \M{D}_0\,\M{m}_0\,\M{D}_0\quad{\rm with}\nonumber\\
\M{m}_0 &=& \nonumber\\ &&\hspace{-1cm}{\rm block~diag}\left({\rm sign}\mu_1,\ldots,{\rm sign}\mu_r, (-1)^{n_1}\sigma_z,\ldots,(-1)^{n_p}\sigma_z \right)\nonumber\\
{\rm and}\nonumber\\
\M{D}_0 &=& {\rm block~diag}(|\mu_1|^{\frac{1}{2}},\ldots,|\mu_r|^{\frac{1}{2}},|\tau_1|^{\frac{1}{2}}\M{1}_2,\ldots,|\tau_p|^{\frac{1}{2}}\M{1}_2).\nonumber\\
\end{eqnarray}

The diagonal matrix $\M{D}_0$ is hermitian and positive, with $\det\M{D}_0\neq 0$, because $\det\M{m}\neq 0$  by assumption. 
Each of the $\sigma_z$ blocks in $\M{m}_0$ in \eqref{m-inertia} has eigenvalues $\pm 1$, and thus, the inertia of $\M{m}$, which equals that of $\M{M}$, is 
\begin{eqnarray}\label{inertia}
i_+ &=& \#{\rm positive~eigenvalues} = p + \#{\rm positive}\, \mu_i's\nonumber\\
i_- &=& \#{\rm negative~eigenvalues} = p + \#{\rm negative}\, \mu_i's\nonumber\\
\end{eqnarray}
(and by assumption, $i_0 =  \#{\rm null~eigenvalues} = 0$).
Therefore, the fact that the given PH matrix $\M{H}$ has $p$ pairs of complex conjugate eigenvalues sets a floor $p$ to the numbers of positive and negative eigenvalues of the metric, and any additional positive or negative eigenvalue of $\M{m}$ originate from the free parameters $\mu_i$.

The metric $\M{m}$ can be expressed in terms of its diagonal form $\M{m}_0$ in \eqref{m-inertia} as 
\begin{equation}\label{m-diagonal1}
\M{m} = {\mathcal U}^\dagger\,\M{D}_0\,\M{m}_0\,\M{D}_0{\mathcal U} = \M{D}_0\, {\mathcal U}^\dagger\,\M{m}_0\,{\mathcal U} \M{D}_0,
\end{equation}
where 
\begin{equation}\label{unitary}
{\mathcal U} = {\rm block~diag}(\underbrace{1,\ldots,1}_{\text{r}},U_1,\ldots,U_p)
\end{equation}
is obtained by setting all $n_s=0$ in \eqref{unitary'}, and where in the last equality in \eqref{m-diagonal1} we have used $[\M{D}_0, {\mathcal U}]=0$. Plugging \eqref{m-diagonal1} in \eqref{Mm} we thus obtain 
\begin{equation}\label{Mm1}
\M{M} = (\M{D}_0\M{S})^\dagger {\mathcal U}^\dagger\,\M{m}_0\,{\mathcal U} (\M{D}_0\M{S}),
\end{equation}
which suggests, borrowing terminology from gauge theory, that $\M{D}_0$ could be ``gauged away" by absorbing it in the diagonalizing matrix $\M{S}$ of $\M{H}$. According to the discussion around Eqs.\eqref{D} and \eqref{eigenvectors}, changing $\M{S}\rightarrow\M{D}_0\M{S}$ does not affect $\M{H}$. Therefore, the data stored in $\M{D}_0$, namely, the absolute values of eigenvalues of $\M{m}$, are completely ``pure gauge". Only the inertia \eqref{inertia} stored in $\M{m}_0$, and the phases $\arg \tau_s$ stored in ${\mathcal U}$, are ``physical" and affect $\M{m}$ and $\M{M}$. Thus, we arrive at the {\em canonical} (or ``unitary gauge") form of our solution of \eqref{intertwining} for the metric $\M{M}$ as
\begin{equation}\label{canonical}  
\M{M} = \M{S}^\dagger {\mathcal U}^\dagger\,\M{m}_0\,{\mathcal U}\M{S}.
\end{equation}

The $r$ signs ${\rm sign}\mu_i = \pm 1$ as well as the $p$ choices of $n_s=0,1$ are independent of each other. Thus, there are $2^{r+p} = 2^{N-p}$ independent possibilities to choose them, or equivalently, $2^{N-p}$ possible matrices $\M{m}_0$. Therefore, the set of all possible $\M{m}_0$'s is essentially $\ZZ_2^{N-p}$. The matrix $\M{m}$ in \eqref{m-solution} is $2\pi$-periodic in each of the phases $\arg\tau_s$. For each choice of $\M{m}_0$, the $p$ phases $0\leq\arg\tau_s <2\pi$ vary independently, thus comprising a $p$-dimensional torus $T^p$. We therefore conclude that the space ${\mathcal M}$ of all solutions of \eqref{intertwining} for the metric $\M{M}$ is isomorphic to $\ZZ_2^{N-p}\otimes T^p$.

To be more precise, the latter should be modded out by one {\em global} $\ZZ_2$ factor corresponding to $\M{M}\rightarrow -\M{M}$, because inverting the sign of the metric does not affect the intertwining relation \eqref{intertwining}. Equivalently, $\M{H}$ and $-\M{H}$ should obviously lead to the same set of metrics. Thus, 
\begin{equation}\label{spaceM}
{\mathcal M} = \ZZ_2^{N-p}\otimes T^p/\ZZ_2^{\rm global}.
\end{equation}

\section{Discussion}\label{Discussion}
Each choice of an element $\M{M}\in{\mathcal M}$ corresponds to an inner product vector space ${\mathcal V}_{\M{M}}$, possibly of indefinite signature, which accommodates the given PH matrix $\M{H}$ as an element of the linear space (with real coefficients) of PH matrices which act on its vectors. Each such PH matrix  
$\M{\Phi}$ can be expressed uniquely \cite{FR-review, FR-B, carlson} as\footnote{This result appeared in the physics literature first in \cite{JK}, in the context of SQH matrices.}  
\begin{equation}\label{observable}
\M{\Phi} = \M{A}\M{M}, \quad \M{A}^\dagger = \M{A}, 
\end{equation}
provided $\det\M{M}\neq 0$, which is what we assume in the first place. 

The special case in which both $\M{A}$ and $\M{M}$ are real symmetric matrices leads to the real pseudo-symmetric matrices constructed in \cite{KA}. On the other hand, a generic diagonalizable real matrix $\M{H}$, whose spectrum automatically consists of either real or pairs of complex conjugate eigenvalues, will be diagonalized by a complex matrix $\M{S}$  as in \eqref{generic}, and is of no exception to the general analysis in this paper. As was observed in \cite{KA}, such an $\M{H}$ fulfills the intertwining relation $\M{H}^T(\M{S}^T\M{S}) = (\M{S}^T\M{S})  \M{H}$, meaning that $\M{H}$ and $\M{H}^T$ are similar. However,  $\M{S}^T\M{S}$ is in general not hermitian and therefore cannot serve as metric. 

For the special case in which $\M{H}$ is SQH, that is, has purely real spectrum of eigenvalues \eqref{spectrum}, we have $r=N$, $p=0$ and \eqref{spaceM} becomes simply $\ZZ_2^N/\ZZ_2^{\rm global}$. The simplest element in this set obviously corresponds to $\M{m}_0 = \M{1}_N$, in which case we obtain from \eqref{canonical} the well-known factorization $\M{M} = \M{S}^\dagger \M{S}$. (Cf. Eq.(13) in \cite{Mostafazadeh1}, or equivalently, Eq.(22) in \cite{Mostafazadeh2}.) 

However, note that even for this special case of an SQH matrix $\M{H}$, there are exponentially many more possibilities for $\M{m}_0$ with all possible sign assignment, all of which lead to vector spaces ${\mathcal V}_{\M{M}}$ of indefinite metric, which accommodate $\M{H}$ as an observable.

As a final remark for this paper, the next step along the lines presented here, is to extend our results to non-diagonalizable matrices $\M{H}$ with a given Jordan form. 

~~~{\em Acknowldegements}
This research was supported by the Israel Science Foundation (ISF) under grant No. 2040/17.


\end{document}